%% file: SPAWC2021_probabilistic_scheduling_in_FEEL_final.tex
\documentclass[10pt, conference, letterpaper]{IEEEtran}

\usepackage{fancyhdr}
\usepackage{epsfig}
\usepackage{threeparttable}
\usepackage{epsf,epsfig}
\usepackage{amsthm}
\usepackage[cmex10]{amsmath}
\usepackage{amssymb, amsfonts}
\usepackage[noadjust]{cite}
\usepackage{dsfont}
\usepackage{subfigure}
\usepackage{color}
\usepackage{soul}
\usepackage{amssymb}
 \usepackage{accents}
  \usepackage{algorithm}
    \usepackage{url}
    \usepackage{framed}

\include{header}

\DeclareMathSizes{10}{9}{5}{3}

\usepackage{mathrsfs}
\usepackage{algorithm}
\usepackage{algorithmic}
\begin{document}	

\title{\huge Accelerating Federated Edge Learning via Optimized Probabilistic Device Scheduling   \vspace{-5pt}}

\author{\IEEEauthorblockN{Maojun Zhang\IEEEauthorrefmark{4},
	     Guangxu Zhu\IEEEauthorrefmark{1}, 
	     Shuai Wang\IEEEauthorrefmark{2}
	     Jiamo Jiang\IEEEauthorrefmark{5}
              Caijun Zhong\IEEEauthorrefmark{4},
              Shuguang Cui\IEEEauthorrefmark{3}\IEEEauthorrefmark{1}
             }
\IEEEauthorblockA{\IEEEauthorrefmark{4} College of information Science and Electronic Engineering, Zhejiang University, Hangzhou, China\\ 
\IEEEauthorblockA{\IEEEauthorrefmark{1}Shenzhen Research Institute of Big Data, Shenzhen, China\\ 
}
\IEEEauthorblockA{\IEEEauthorrefmark{2} Southern University of Science and Technology, Shenzhen 518055, China}
\IEEEauthorblockA{\IEEEauthorrefmark{5} China Academy of Information and Communications Technology, Beijing, China\\ 
}
\IEEEauthorblockA{\IEEEauthorrefmark{3}FNii and SSE, The Chinese University of Hong Kong (Shenzhen), Shenzhen, China\\ 
}
\IEEEauthorblockA{Email: zhmj@zju.edu.cn, gxzhu@sribd.cn, wangs3@sustech.edu.cn, \\ jiangjiamo@caict.ac.cn, caijunzhong@zju.edu.cn, shuguangcui@cuhk.edu.cn}
}
\vspace{-5mm}}

\maketitle


\begin{abstract}
The popular \emph{federated edge learning} (FEEL)  framework allows  privacy-preserving collaborative model training via frequent learning-updates exchange between edge devices and server. 
Due to the constrained bandwidth, only a subset of devices can upload their updates at each communication round. This has led to an active research area in FEEL studying the optimal device scheduling policy for minimizing communication time. However, owing to the difficulty in quantifying the exact communication time, prior work in this area can only tackle the problem  partially by considering either the communication rounds or per-round latency, while the total communication time is determined by both metrics. To close this gap, we make the first attempt in this paper to formulate and solve the communication time minimization problem. 
We first derive a tight bound to approximate the communication time through cross-disciplinary effort involving both  learning theory for convergence analysis and communication theory for per-round latency analysis. 
Building on the analytical result, an optimized probabilistic scheduling policy is derived in closed-form by solving the approximate communication time minimization problem. 
It is found that the optimized policy gradually turns its priority from suppressing the remaining communication rounds to reducing per-round latency as the training process evolves. 
The effectiveness of the proposed scheme is demonstrated via a use case on collaborative 3D objective detection in autonomous driving.
\end{abstract}

\vspace{-1mm}
\section{introduction}
\vspace{-1mm}
In the pursuit of truly brain-like intelligence, \emph{federated edge learning} (FEEL) has emerged as a popular framework for collaborative model training over wireless networks \cite{mcmahan2017communication}, which allows full exploitation of the rich distributed data at the edge devices without compromising the data privacy. This is achieved by distributing the learning task across edge devices and letting each of them to upload the computed learning updates (e.g., local gradients or models) instead of the raw data to the edge server, where local updates are aggregated to yield an updated global model for initiating the next round local model training.  When the edge devices participating in the learning process share the same wireless medium to convey local updates to the edge server, the uncertain wireless environment and limited radio resources can cause severe congestion over the air interface, resulting in a communication bottleneck for FEEL. This thus prompts an active research area focusing on developing novel communication techniques (e.g., resource allocation, multiple access, source coding) for communication-efficient FEEL \cite{zhu2020toward}. 
	
	Among others, device scheduling in FEEL is an important issue that received increasing research interests recently while yet to be fully addressed. The research issue is prompted by the fact that, due to the constrained bandwidth, only a subset of devices can be selected to convey their updates at each communication round in FEEL.  Simple uniform random scheduling is first considered in the seminal work \cite{mcmahan2017communication} to determine the subset of devices for update uploading. Despite its simplicity, uniform random scheduling shows decent convergence performance as observed empirically in \cite{mcmahan2017communication}. Subsequently, authors in \cite{yang2019scheduling} compared three heuristic scheduling policies including uniform random scheduling, round robin, and proportional fair, where the convergence behavior of the three policies are theoretically characterized. 	
Later on, a joint device scheduling and resource allocation policy was proposed in \cite{wang2019adaptive} to maximize the model accuracy within a given total training time budget.
Realizing that the device scheduling policy can crucially affect the total communication time of FEEL, recent research effort in this field focused on studying the optimal scheduling policy that minimizes the communication time. 
However, due to the difficulty in quantifying the exact communication time, prior work in this area can only tackle the problem  partially by considering either the communication rounds or per-round latency, while the total communication time is determined by both metrics. In terms of communication-round minimization, authors in \cite{rizk2020optimal} proposed an \emph{importance-aware scheduling policy} where the devices with larger gradient norms are deemed to be more important and thus assigned a higher probability to be scheduled. Similar policies are also proposed independently in  parallel works \cite{chen2020optimal,amiri2020convergence,chen2020convergence}.  In terms of per-round latency minimization, authors in \cite{zhu2019broadband} proposed a \emph{channel aware scheduling policy}, where devices with stronger channels are scheduled with higher frequency than the those with weak channels.    
Most recently, authors in \cite{ren2020scheduling} proposed an indirect approach to reduce the total communication time by minimizing a heuristic objective function which is constructed to be a weighted sum of the gradient norm and per-round communication time. The weighted factor therein crucially affects the performance of the derived solution, and thus need to be fine-tuned in an offline manner, which may not be feasible in practical FEEL systems.

In this paper, we made the first attempt to directly formulate and solve the communication time minimization problem. We first derive a tight bound to approximate the communication time through cross-disciplinary effort involving both learning theory for convergence analysis and communication theory for per-round latency analysis. 
Building on the analytical result, an optimized probabilistic scheduling policy is derived in closed-form by solving the approximate problem. 
It is found that the optimized policy gradually turns its priority from suppressing the remaining communication rounds to reducing per-round latency as the training process evolves. 
The effectiveness of the proposed scheme is demonstrated via a use case on collaborative 3D objective detection in autonomous driving.



\vspace{-1mm}
\section{Learning and Communication Models}
	We consider a FEEL system, which consists of one edge server and $M$ different kinds of edge devices. 
	The device set can be denoted by $\mathcal{M}=\{1,2,3,...,M\}$. 
	The local dataset is denoted by ${\cal D}_m$. 
	The size of ${\cal D}_m$ is denoted by $n_m$. 
	We let $n =\sum_{m=1}^{M}n_m$ denote the total dataset size. 
	Due to the constrained bandwidth, only a subset of devices can be scheduled to participate in the global model updating process in each round. A scheduler determining the participating devices at each  round need to be judiciously designed at each communication round for such a system. 
\vspace{-1mm}
\subsection{Learning Model}\label{subsec:learning model}
The learning process is to minimize the global loss function in a distributed manner. Particularly, the global loss function on the entire distributed dataset is defined below:
	\begin{align}\label{eq:loss_function_definition}  \qquad
		L(\mathbf{w}^{(t)}) = \frac{1}{\sum_{m=1}^{M}{n_m}} \sum_{m=1}^{M} \sum_{(\bx_j, y_j) \in {\cal D}_m} f(\bw^{(t)}, \bx_j, y_j),
	\end{align}
	where $\mathbf{w}^{(t)}$ denotes the global model at the round $t$,
	 $f(\bw^{(t)}, \bx_j, y_j)$ denotes the sample loss quantifying the prediction error of $\mathbf{w}^{(t)}$ on the training data $\bx_j$ with respect to the true label $y_j$.
	At each communication round, FEEL repeats the following procedure until the global model converges:
		\begin{itemize}
		\item {{\bf Global Model Broadcasting}: The edge server broadcasts current global model $\mathbf{w}^{(t)}$ to each device.}
		\item{{\bf Local Model Training}: Each device runs the \emph{stochastic gradient decent} (SGD) algorithm using its local dataset and the latest global model  $\mathbf{w}^{(t)}$, and generates a local gradient estimate $\mathbf{g}^{(t)}_{m}$.}  
		\item{{\bf Probabilistic Device Scheduling}: Each device is assigned a certain probability, denoted by $p_m^{(t)}$ for device $m$ to be designed in the sequel, for being scheduled for update uploading. The scheduled devices are decided by sampling based on the probability distribution.}
		\item{{\bf Local Gradient Uploading}: Each scheduled device  transmits a scaled version of the local gradient estimate to the edge server, which is given by $\widehat{\mathbf{g}}^{(t)}_m=\frac{n_m}{n\cdot p_m^{(t)}}\mathbf{g}_m^{(t)}$.\footnote{The scaling factor $\frac{n_m}{n\cdot p_m^{(t)}}$ is needed to ensure an unbiased gradient estimate at the edge server as explained in \cite{ren2020scheduling}.}}
		\item{{\bf Global Model Updating}: The edge server aggregates the uploaded local gradients, and then updates the global model by  $\mathbf{w}^{(t+1)} = \mathbf{w}^{(t)}-\eta^{(t)}\widehat{\mathbf{g}}^{(t)}$.}
	\end{itemize} 

\vspace{-2mm}
\subsection{Communication Model} \label{subsec:communication model}
Each scheduled device is assigned a dedicated sub-channel of bandwidth $B$ for update uploading. 
For device $m$, the uploading time $T_{{\sf U},m}^{(t)}$ at round $t$ is given by
	\begin{align}\label{eq:upload time of user m}
	T_{{\sf U},m}^{(t)} = \frac{qd}{B\log_{2}{(1+\gamma_{m}^{(t)})}},
	\end{align}
	where $q$ is the number of bits needed for transmitting one gradient parameter, $d$ is the total number of gradient parameters, $\gamma_{m}^{(t)}$ denotes the \emph{signal-to-noise ratio} (SNR) of the transmitted signal, which is defined as $\gamma_{m}^{(t)}=\frac{P_{m}|{h}_{m}^{(t)}|^2}{N_0}$, 
	with $P_m$ being the transmit power, $N_0$ the noise power, and $|{h}_m^{(t)}|^2$ the channel gain. 
	The channel coefficients $\{{h}_m^{(t)}\}$ are  assumed to be \emph{independent and identically distributed} (i.i.d.) over time, 
	following Rayleigh distribution, i.e., ${h}_m^{(t)} \sim {\cal CN}(0, \sigma_m^2)$, where the channel variance varies in devices due to the heterogeneous path losses encountered. 

\vspace{-2mm}
\section{Problem Formulation}\label{sec:problem formulation}
\vspace{-1mm}

Consider the probabilistic scheduling scheme described in Section \ref{subsec:learning model}. For each communication round, a scheduling problem is formulated to optimize the scheduling probability distribution for minimizing the remaining communication time. The problem instance for round $t$ is given by
\begin{align}\label{problem:first globle optimize problem}
	\min_{p_{1}^{(t)},..,p_{M}^{(t)}} T^{(t)}=\sum_{i=t}^{N_t} T_{\sf C}^{(i)}, \quad  \mathbf{s.t.} \sum_{m=1}^{M} p_m^{(t)} =1,
\end{align}
where $T^{(t)}$ and $N_t$ denote the remaining communication time and  the remaining number of communication rounds when sitting at round $t$, $T_{\sf C}^{(i)}$ denotes the one-round communication time at round $i$, with $t\leq i \leq N_t$. 
Particularly,  the global model $\mathbf{w}^{(N)}$ at round $N$ is deemed converged if  \emph{$\epsilon$-accuracy} is achieved as defined below:  
\begin{align}\label{eq:sign of global model convergence}
	|L(\mathbf{w}^{(N)})-L(\mathbf{w}^*)|\leq \epsilon.
\end{align}
According to the learning procedure in Section \ref{subsec:learning model}, the communication time at round $i$ consists of  two parts, i.e.,
\begin{align}\label{one_round_time}
	T_{\sf C}^{(i)} =T^{(i)}_{\sf B}+T^{(i)}_{\sf U},
\end{align}
where $T_{\sf B}^{(i)}$ and $T_{\sf U}^{(i)}$ represent the global model broadcasting time and the local gradient uploading time at round $i$, 
respectively. 
Since $T_{\sf B}^{(i)}$  is  independent with the scheduling decision, the optimization problem in \eqref{problem:first globle optimize problem} thus reduces to
\begin{align}\label{problem:first globle optimize problem 2}
	\min_{p_{1}^{(t)},..,p_{M}^{(t)}} \sum_{i=t}^{N_t} T_{\sf U}^{(i)}, \quad  \mathbf{s.t.} \sum_{m=1}^{M} p_m^{(t)} =1.
\end{align}
However, the objective above is not tractable for evaluation in practice as the involved $N_t$ and $\{T_{\sf U}^{(i)}\}_{i=t+1}^{N_t}$ require non-causal future information (e.g., the gradient estimates, channel state information at future rounds) for exact calculation. 
The difficulty can be tackled by formulating an approximate problem using the \emph{look-ahead model} \cite{powell2014clearing}, 
which is widely used in stochastic optimization society, as follows
	\begin{align}\label{problem:second globle optimize problem}
	\mathscr{P}_{1}:&\min_{p_{1}^{(t)},..,p_{M}^{(t)}} T_{\sf U}^{(t)}+N_{t+1}^{\mathbb{E}}T_{\sf U}^{\mathbb{E}},\quad  \mathbf{s.t.} \sum_{m=1}^{M} p_m^{(t)} =1,
	\end{align}  
	where 
	$N_{t+1}^{\mathbb{E}}$  denotes the expected remaining communication rounds after round-$t$'s update, 
	and 
	$T_{\sf U}^{\mathbb{E}}$ denotes the expected communication time of future round. 
	Compared with in \eqref{problem:first globle optimize problem 2}, the one in problem $\mathscr{P}_{1}$ requires only causal information and thus can be practically evaluated as shown shortly.

\section{Scheduling Optimization}\label{sec:scheduling design}
In this section, we first derive the three key components in the objective  of problem $\mathscr{P}_1$, i.e., $T_{\sf U}^{(t)}$, $N_{t+1}^{\mathbb{E}}$, and $T_{\sf U}^{\mathbb{E}}$, as functions of scheduling probability distribution $\{p_m^{(t)}\}$. Based on the analytical result, the optimized $\{p_m^{(t)}\}$ are then attained in closed-form by solving problem $\mathscr{P}_1$.

\subsection{Communication Time Analysis for FEEL} \label{subsec:Training time Analysis for FEEL}
\subsubsection{Expected Remaining Communication Rounds}
Since the exact $N_{t+1}^{\mathbb{E}}$ is not tractable to derive, we thus resort to deriving an upper bound as the approximation of $N_{t+1}^{\mathbb{E}}$. To this end, the following standard assumptions on the loss function are made.   
\begin{assumption}The loss function $L(\cdot)$ is $\ell$-smooth, i.e., $\forall\ \mathbf{u}\ and\ \mathbf{v},L(\mathbf{u}) \leq L(\mathbf{v})+\nabla L(\mathbf{v})^{\top}(\mathbf{u}-\mathbf{v})+\frac{\ell}{2}\|\mathbf{u}-\mathbf{v}\|^{2}$.
\end{assumption}
\begin{assumption} The loss function $L(\cdot)$ is $\mu$-strongly-convex, i.e., $\forall\ \mathbf{u}\ and\ \mathbf{v},L(\mathbf{u}) \geq L(\mathbf{v})+\nabla L(\mathbf{v})^{\top}(\mathbf{u}-\mathbf{v})+\frac{\mu}{2}\|\mathbf{u}-\mathbf{v}\|^{2}$.
\end{assumption}
Then the upper bound of $N_{t+1}^{\mathbb{E}}$ can be derived as a function of $\{p_m^{(t)}\}$ as follows.
	\begin{proposition}\label{theo:communication upper bound}
		\emph{For FEEL with probabilistic scheduling, the upper bound of remaining communication rounds is given by}
		\begin{align}\label{eq:communication round still needed}
		N_{t+1}^{\mathbb{E}} \leq \frac{\ell(t+1+\nu)\left(\eta^{(t)}\right)^2}{2\epsilon} \sum_{m=1}^{M} {\left(\frac{n_m}{n}\right)^2\frac{\|{\mathbf{g}}_{m}^{(t)}\|^2}{p_{m}^{(t)}}} + C^{(t+1)},
		\end{align}
	\end{proposition}
\noindent where a diminishing stepsize $\eta^{(t)}=\frac{\chi}{t+\nu}$ is adopted with constants $\chi$ and $\nu$;   $C^{(t+1)}=\frac{\ell  \chi^2 \left(G^{(t+1)}\right)^2}{2\epsilon(2\mu\chi-1)}+\frac{\left(t+\nu+1\right)\left(\frac{1}{2\mu}-\eta^{(t)}\right)}{\epsilon}\|{\mathbf{g}^{(t)}}\|^2-\nu-t-1$ is a constant term independent of the scheduling probability distribution; 
$G^{(t+1)}$ is the maximum expected uploaded gradient norm from the $(t+1)$-th round to the last round. 
\begin{proof}
	See Appendix A.
\end{proof}
\begin{remark}
\emph{
With Theorem \ref{theo:communication upper bound} at hand, by Cauchy inequality, it can be proved that the scheduling probability distribution  minimizing  the remaining communication rounds is proportional to the product of local dataset size and gradient norm, i.e., $p_m^{(t)} \propto n_m \| \mathbf{g}_m^{(t)}\|$. This aligns with the \emph{importance-aware scheduling} proposed in \cite{chen2020optimal} where the product $n_m \| \mathbf{g}_m^{(t)}\|$ defines the update importance of device $m$. 
}
\end{remark}

\subsubsection{Uploading Time at Current Round}
Due to the probabilistic scheduling, $T_{\sf U}^{(t)}$ can only be estimated by its expectation over the scheduling probability distribution, yielding
	\begin{align}\label{current round expected communication time}
		T_{\sf U}^{(t)}=\sum_{m=1}^{M}p_{m}^{(t)} T_{{\sf U},m}^{(t)},
	\end{align}
	where $T_{{\sf U},m}^{(t)}$ has been defined in \eqref{eq:upload time of user m}, which can be exactly computed using channel state information at the current round.

	\begin{proposition}\label{prop:2}
		\emph{For FEEL with probabilistic scheduling, the uploading time of current round is given by}
		\begin{align}\label{expected time in t-th round final form}
			T_{{\sf U}}^{(t)}=\sum_{m=1}^{M}{p_m^{(t)}}T_{{\sf U},m}^{(t)}=\sum_{m=1}^{M}{p_m^{(t)}}\frac{q d}{BR_{m}^{(t)}},
		\end{align} 
\emph{where we have defined $R_{m}^{(t)}=\log _{2}\left(1+\gamma_{m}^{(t)}\right)$.}		
	\end{proposition}
\begin{remark}\emph{
It can be observed that for minimizing the current round uploading time, the optimal solution is to assign the device with the strongest channel (or largest $R_m^{(t)}$) with probability one. This reduces to a deterministic scheduling policy referred to be \emph{channel-aware scheduling}.
}
\end{remark}

\subsubsection{Expected Future Round Communication Time} Due to the absence of the gradient information, channel state information, and scheduling decision at future rounds, the exact communication time at future founds is not tractable to compute and thus we resort to its approximation by its expectation, which is taken over both the random scheduling and random channel distributions. Then,  $\forall i > t$, with $t$ being the index of current round, we have
\begin{align}
T_{{\sf U}}^{\mathbb{E}}=\sum_{k=1}^{M}\frac{qdn_{m}}{Bn}\mathbb{E}_{h_m^{(i)}}\left\{\frac{1}{R_{m}^{(i)}}\right\},
\end{align}
	where the future rounds' scheduling probability is unknown and assumed to be proportional to the size of local dataset, i.e., $p_m^{(i)} = \frac{n_m}{n}$, $\forall i > t$.   

	To ensure efficient transmission, our scheduling policy assign non-zero scheduling probability to those devices with a channel gain larger than a pre-defined threshold $g_{th}$. Only those with $|h_m^{(i)}|^2 \geq g_{th}$ have chance to participate to the FEEL. 
	Then $\mathcal{Q}_m \triangleq\mathbb{E}_{h_m^{(i)}}\left\{\frac{1}{R_{m}^{(i)}}\right\}$ can be computed by
	\begin{align}\label{calculate expectation of fraction of bit rate}
	\mathbb{E}_{h_m^{(i)}}\left\{\frac{1}{R_{m}^{(i)}}\right\} = \int_{g_{th}}^{+\infty}\frac{1}{\sigma_m^2\log_2{\left(1+\frac{P_mz}{N_0}\right)}}e^{-\frac{z}{\sigma_m^2}}dz.
	\end{align} 
	\begin{proposition}
	\emph{For FEEL with probabilistic scheduling, the expected time in future rounds is given by
	}
		\begin{align}\label{expected time in one round final form}
			T_{{\sf U}}^{\mathbb{E}}=\sum_{m=1}^{M}\frac{q dn_{m}\mathcal{Q}_m}{nB}.
		\end{align} 
	\end{proposition}
\subsection{Scheduling Optimization for FEEL}
With the derived $T_{\sf U}^{(t)}$, $N_{t+1}^{\mathbb{E}}$, and $T_{\sf U}^{\mathbb{E}}$ at hand, the problem $\mathscr{P}_1$ can be explicitly written below, where those constant terms independent of the scheduling probability distribution is dropped for simplifying the expression without harming the problem equivalence. 	
	\begin{align}
	\mathscr{P}_2:&\min_{p_{{1}}^{(t)},p_{{2}}^{(t)},...p_{M}^{(t)}}A(t)\left(\eta^{(t)}\right)^2T_{\sf U}^{\mathbb{E}}\sum_{m=1}^{M}\left(\frac{n_m}{n}\right)^2{\frac{\|{\mathbf{g}}_{m}^{(t)}\|^2}{p_{m}^{(t)}}}+{p_m^{(t)}}T_{{\sf U},m}^{(t)}\notag \\ 
	&\mathbf{s.t.}\sum_{m=1}^{M}p_{m}^{(t)}=1,\notag
	\end{align}
	where we have defined $A(t)=\frac{\ell(t+1+\nu)}{2\epsilon}$. 
	Next, the problem $\mathscr{P}_2$ can be solved below.

	\begin{proposition} \label{theo:probability allocation}
		\emph{For FEEL with probabilistic scheduling, the optimized scheduling probability distribution for minimizing the remaining communication time is given by}
		\begin{align}\notag 
			p{_{m}^{(t)*}} \!=\! \rho_t  \frac{\frac{n_m}{n}\|\mathbf{g}{_{m}^{(t)}}\|}{\sqrt{\frac{qd}{BR_{m}^{(t)}} \!+\!\lambda^*}}, \text{with} \; \rho_t \!=\!  \sqrt{\frac{\ell\left(t \!+ \! 1 \! + \! \nu\right) \chi^2}{2(t+\nu)^2\epsilon} \sum_{m=1}^{M}\frac{qdn_{m}\mathcal{Q}_m}{nB}} 
		\end{align}
\emph{where 
 $\lambda^*$ is the Lagrange multiplier satisfying $\sum_{m=1}^{M}p_m^{(t)*} = 1$, and can be attained via bisection search.}
	\end{proposition}
	\proof It can be easily proved that
	$\mathscr{P}_2$ is a convex problem with respect to the scheduling probability distribution $\{p_m^{(t)}\}$. Thus, we can optimally solve this problem using the \emph{Karush-Kuhn-Tucker} (KKT) conditions, yielding
	\begin{align}\label{global communication time optimization problem_solution}
	p{_{m}^{(t)*}} = \frac{n_m}{n}\sqrt{\frac{\ell(t+1+\nu) \left(\eta^{(t)}\right)^2T_{\sf U}^{\mathbb{E}}\|\mathbf{g}{_{m}^{(t)}}\|^2}{2\epsilon\left(T{_{{\sf U},m}^{(t)}+\lambda^*}\right)}},
	\end{align}
		Further substituting $T_{\sf U}^{\mathbb{E}}$ and $T_{{\sf U},m}^{(t)}$  derived respectively in (\ref{expected time in t-th round final form}) and (\ref{eq:upload time of user m}), and $\eta^{(t)}=\frac{\chi}{t+\nu}$ into (\ref{global communication time optimization problem_solution}), gives the desired result.
	\endproof
	
	\begin{remark} 
	\emph{
	As observed in Proposition \ref{theo:probability allocation}, 
	the optimized scheduling probability is governed by three terms, i.e., the one related to gradient importance $\left(\frac{n_m}{n}\|\mathbf{g}{_{m}^{(t)}}\|\right)$, the one related to 
	transmission rate $\left( 1/\sqrt{\frac{qd}{BR_{m}^{(t)}}+\lambda^*}\right)$, and a scalar factor $\rho_t$ regulating the tradeoff between them. 
	This suggests that the optimized scheme tries to balance the consideration between the remaining communication rounds (related to gradient importance) and the uploading time in current round (related to transmission rate). Since $\rho_t$ is a decreasing function w.r.t. $t$, it can be concluded that
the optimized scheme weights more on suppressing the remaining communication rounds at the early training stage while gradually biases for reducing one-round uploading time as the training process evolves.		
	}

	\end{remark}
	\vspace{-2mm}
\section{Experimental Results}\label{simulation}
\vspace{-1mm}

\begin{figure}[tt]
	\centering
	\includegraphics[width=8.5cm]{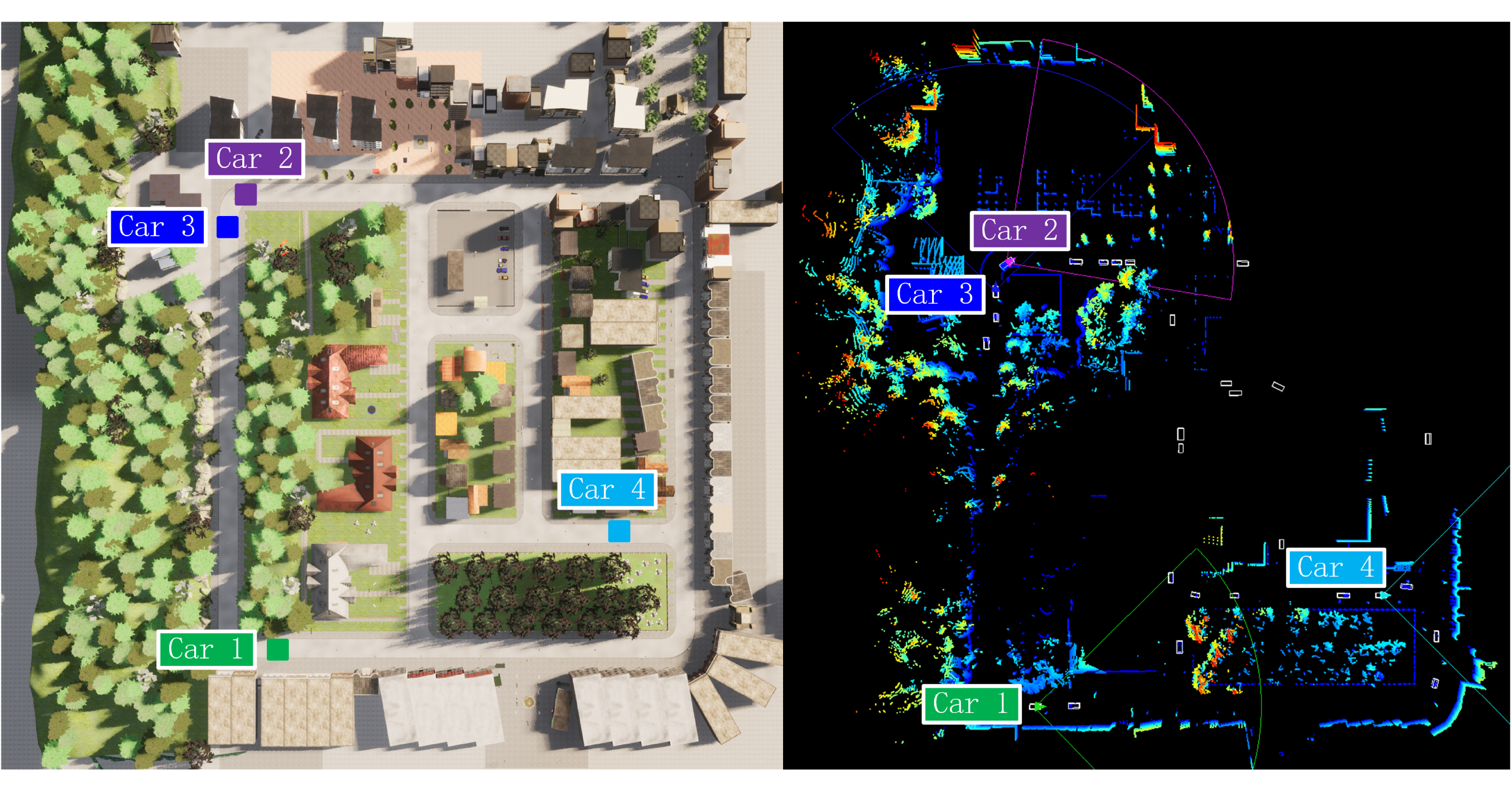}
	\caption{Global bird view and image of one frame}
	\vspace{-3mm}
	\label{Fig:AD_scenario_show}
\end{figure}

\begin{figure}[tt]
	\centering
	\hspace{-10mm}
	\subfigcapskip=-10pt
		\quad

	\subfigure[Early stage, (time=6000s)]{
		\label{Fig:AD_acc_plot:a} 
		\includegraphics[width=4.2cm]{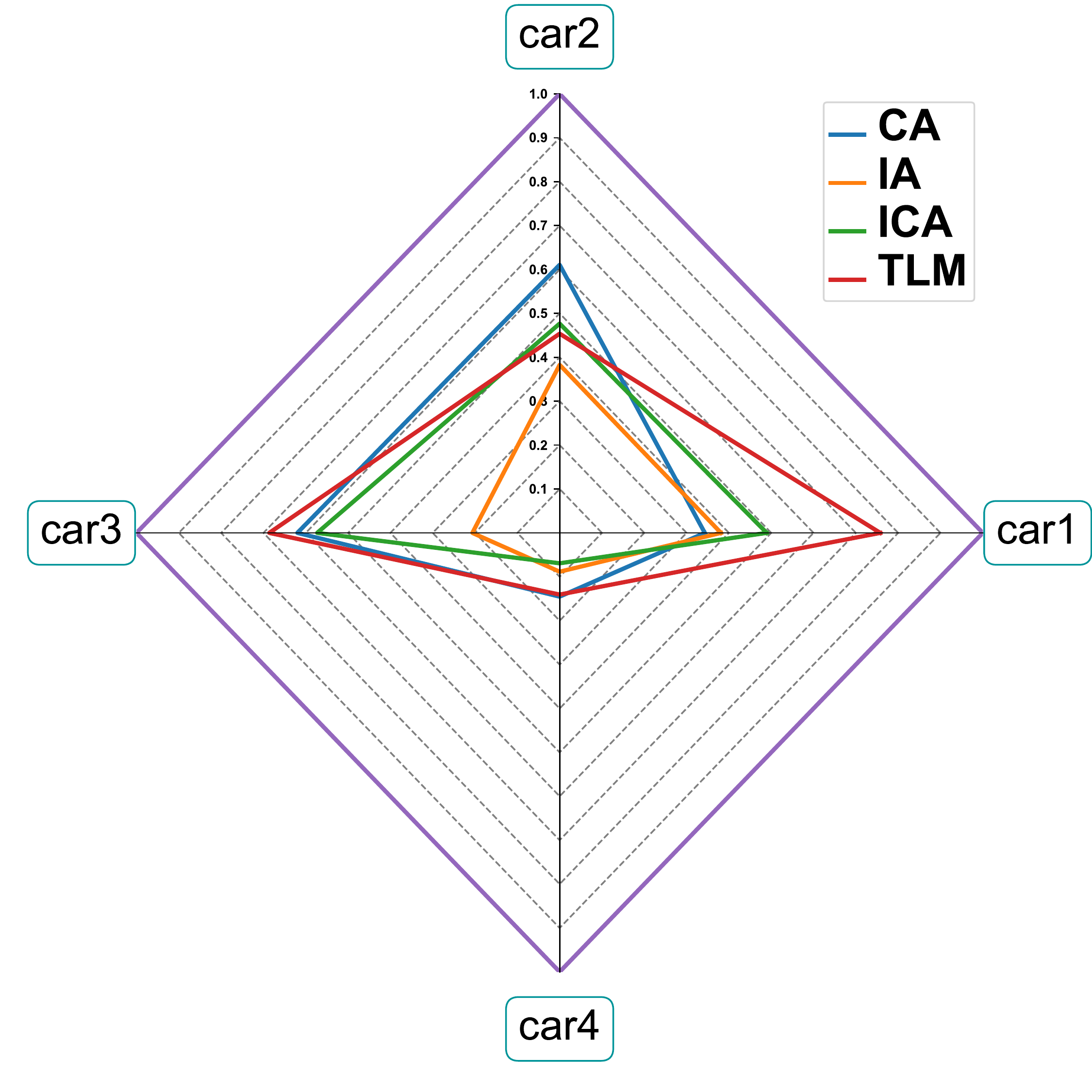}}
	\subfigure[Late stage, (time=14000s)]{
		\label{Fig:AD_acc_plot:c} 
		\includegraphics[width=4.2cm]{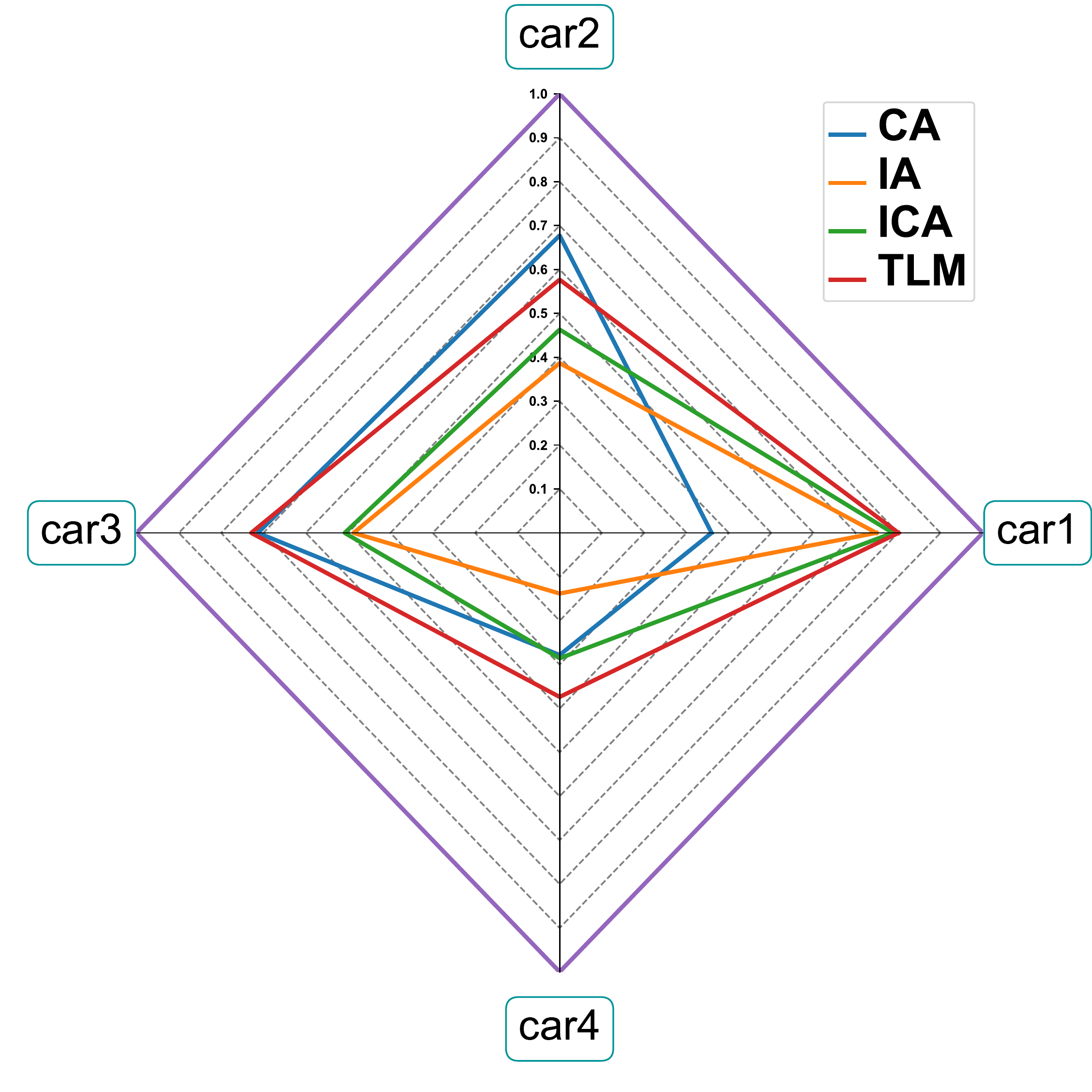}}
	\caption{Test accuracy in different training stages.}
	\label{Fig:AD_acc_plot}
\end{figure}

\begin{figure*}[tt]
	\centering
	\hspace{-10mm}
	\subfigcapskip=0pt
		\quad

	\subfigure[CA]{
		\label{Fig:AD_box_plot:a} 
		\includegraphics[width=0.23\linewidth]{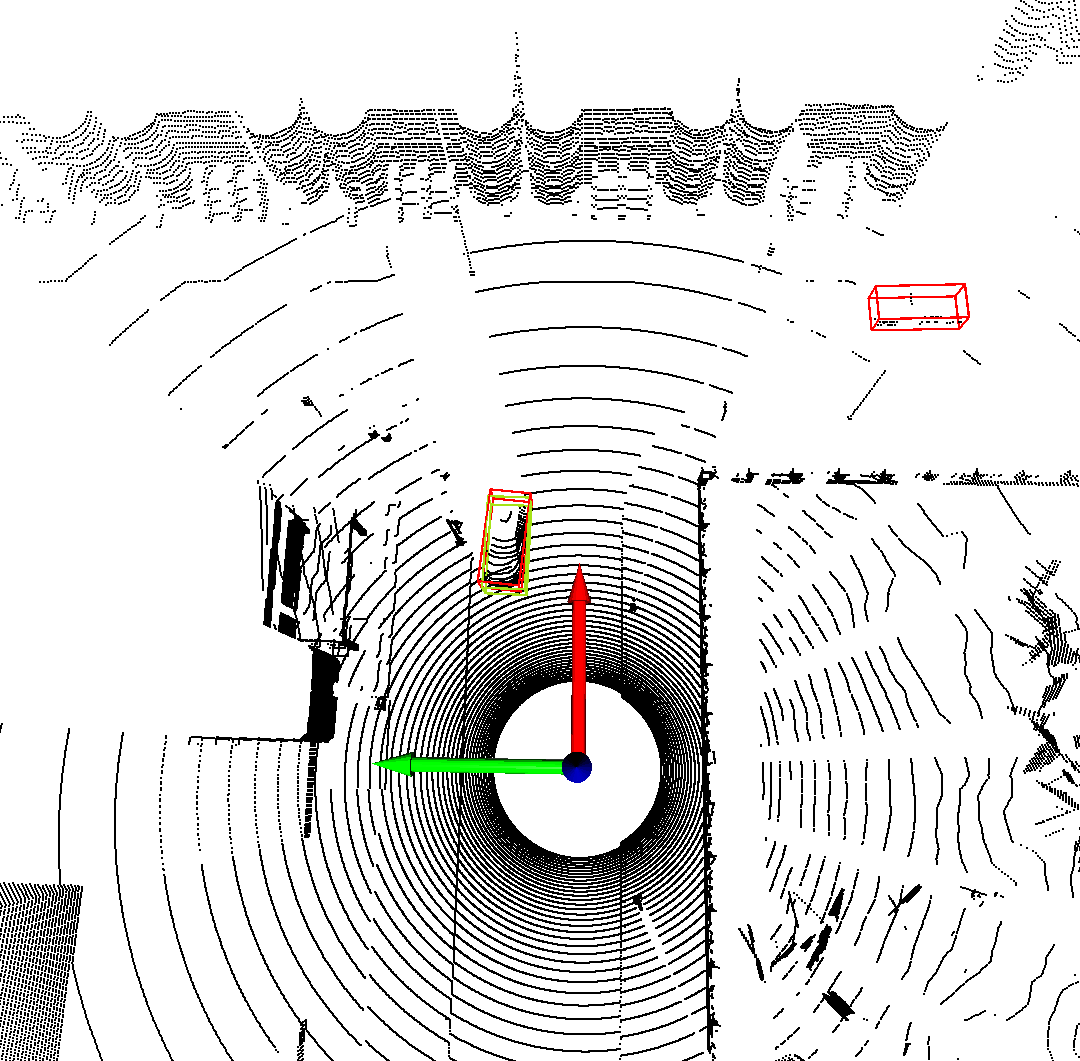}}
	\subfigure[IA]{
		\label{Fig:AD_box_plot:b} 
		\includegraphics[width=0.23\linewidth]{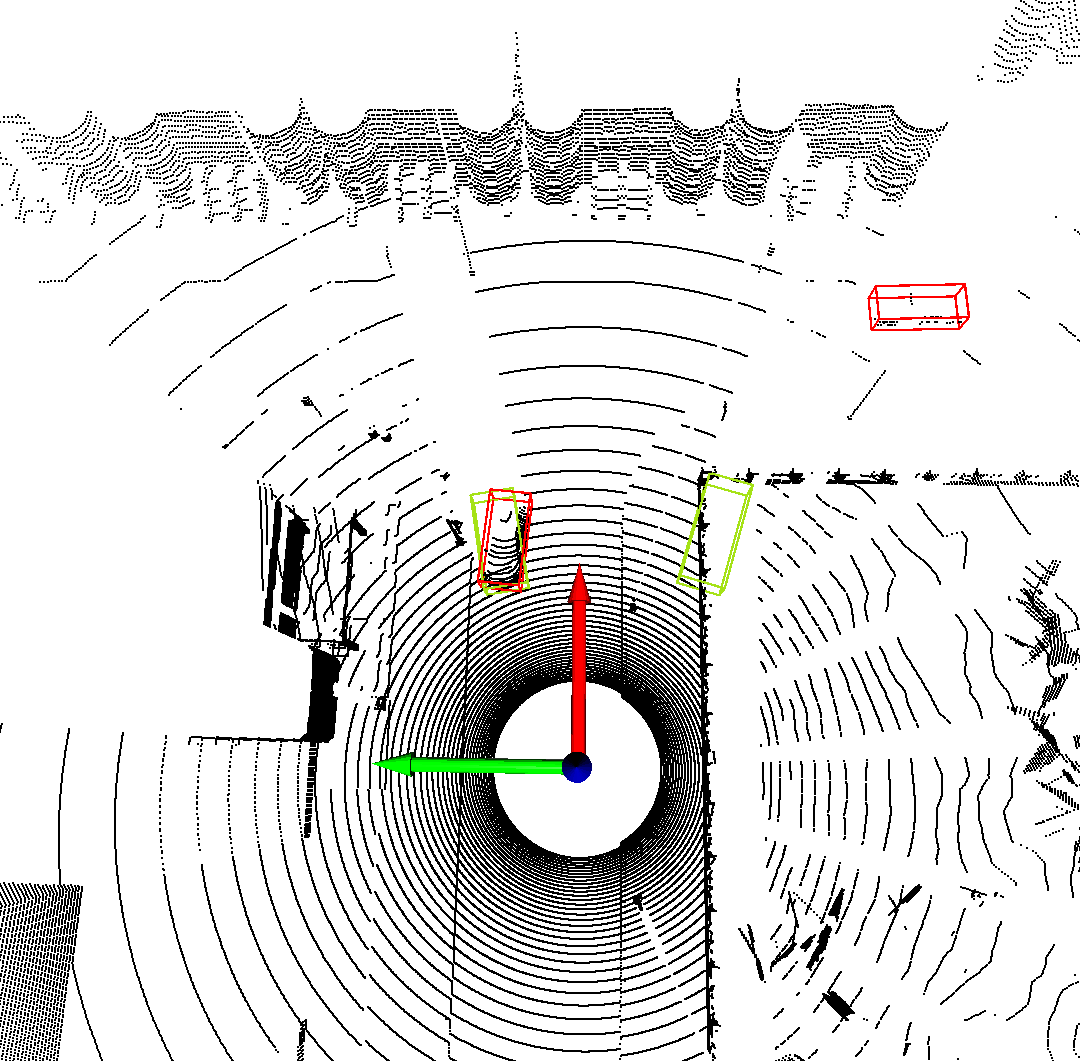}}
	\subfigure[ICA]{
		\label{Fig:AD_box_plot:c} 
		\includegraphics[width=0.23\linewidth]{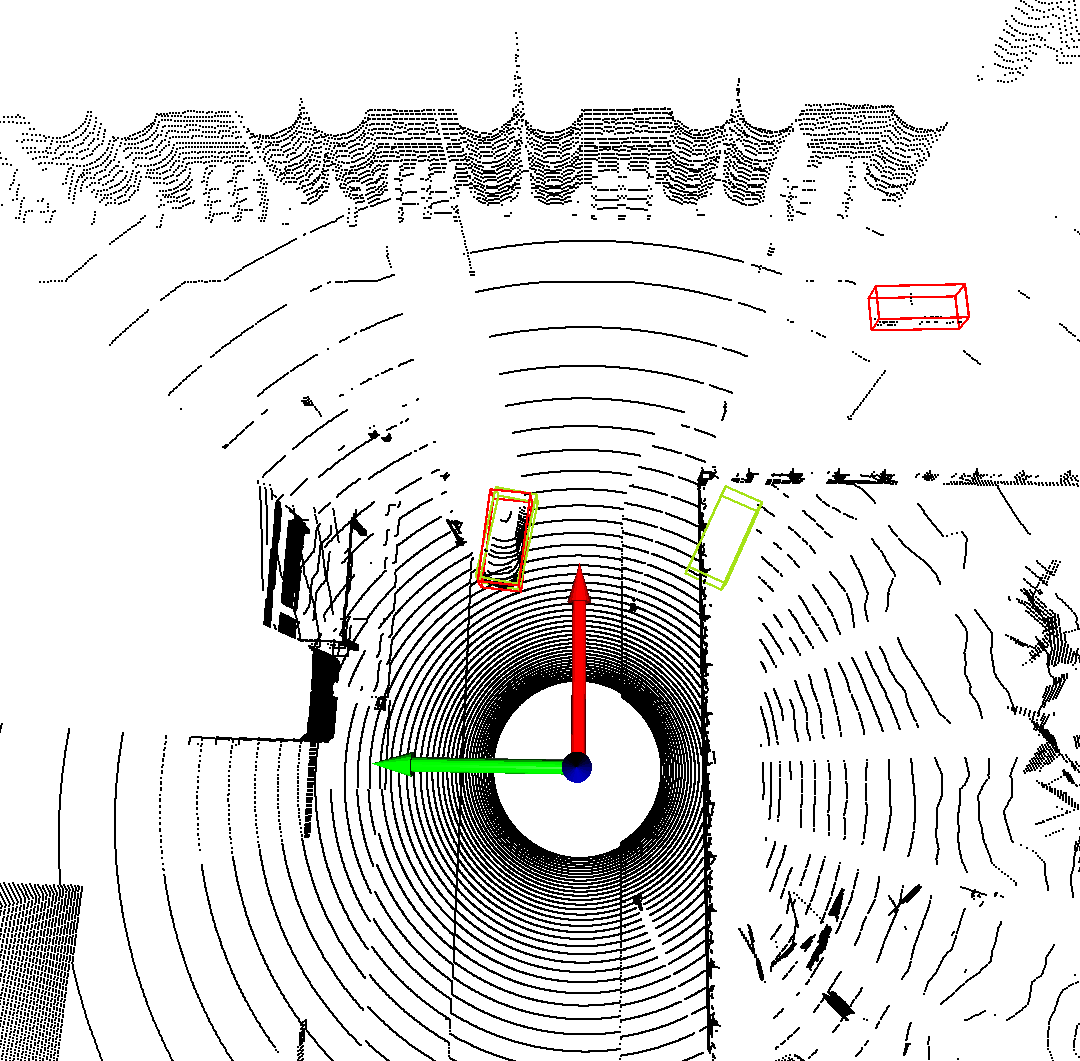}}
	\subfigure[CTM]{
		\label{Fig:AD_box_plot:d} 
		\includegraphics[width=0.23\linewidth]{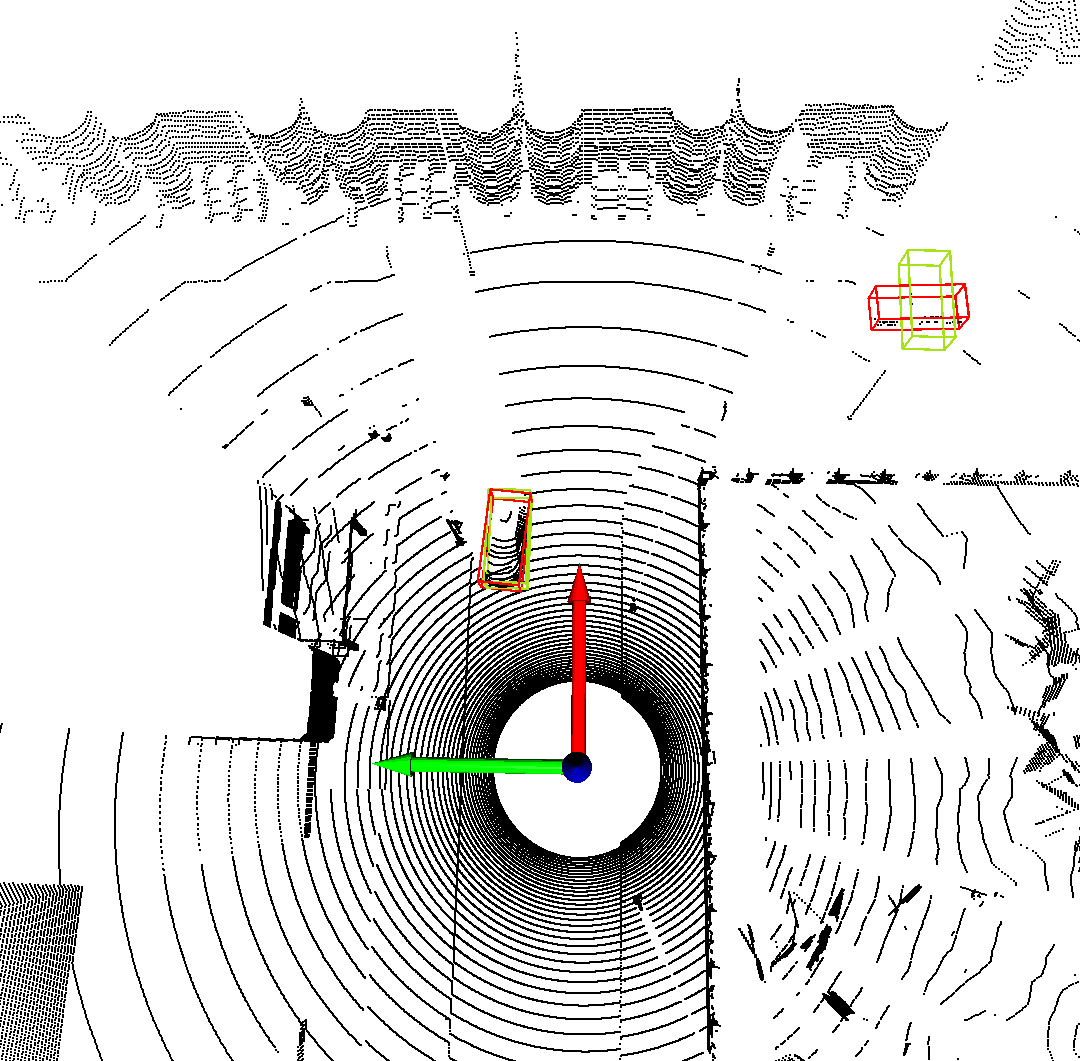}}
	\caption{Performance comparison of 3D object detection for different scheduling schemes.}
	\label{Fig:AD_box_plot}
\end{figure*}

\textbf{Unreal-engine autonomous driving simulator}. Recently, there is a strong demand for implementation of FEEL in autonomous driving vehicle perception to alleviate the communication overhead due to raw data transmission and data privacy issues. 
Unfortunately, physical-world implementation of autonomous driving is hindered by the infrastructure costs and the difficulties of testing in dangerous scenarios. 
\emph{Car Learning to Act} (CARLA) \cite{dosovitskiy2017carla} is a widely-accepted unreal-engine driven benchmark system that provides complex urban driving scenarios and high-quality 3D rendering  such that FEEL-based vehicle perception can be prototyped in virtual-reality.
In this section, all the training and testing procedures are implemented in CARLA.

\textbf{Dataset}. We employ CARLA to generate $32$ vehicles in the ``Town02'' map, among which, $4$ of them are autonomous driving vehicles that can generate the point cloud data at a frequency of $20$ frames/s.
The entire dataset consists of $2800$ frames in total with $700$ frames at each vehicle, where $200$ frames are used for training at each vehicle and $500$ frames are used for inference and testing.
The left hand side of Fig.~1 illustrates the bird eye view of the simulated world and the locations of all vehicles; the right hand side of Fig.~1 illustrates the non-i.i.d. distributed point-cloud data.

\textbf{Model}. The \emph{sparsely embedded convolutional detection} (SECOND) neural network proposed in \cite{yan2018second} is adopted for 3D object detection.
The raw data generated from CARLA is processed into KITTI formats following the procedure in \cite{zhang2021distributed}.
Each round of local training randomly selects $16$ frames and the fraction decay learning rate is adopted.
The federated model training is implemented by PyTorch using python 3.8 on a Linux server with four NVIDIA RTX 3090 GPUs.

\textbf{Communication settings}. 
The FEEL between $1$ edge server and the $4$ autonomous vehicles is executed during vehicle charging. 
The distance $\omega$ in km between the edge server and any vehicle is uniformly distributed between $0.3$ and $0.7$, and the corresponding path loss is given by $128.1+37.6\,\mathrm{log}_{10}(\omega)$ in dB.
The bandwidth, the noise power density, the transmit power of each vehicle, and 
the number of bits for representing each element in gradients are set to be $1\,$MHz, $174\,$dBm/Hz, $24\,$dBm, and $16$, respectively. 

\textbf{Benchmarks}. We compare the proposed \emph{communication time minimization} (CTM) scheme with three benchmark schemes including 1) the \emph{importance-aware} (IA) scheme \cite{rizk2020optimal}; 2) the \emph{channel-aware} (CA) scheme, which schedules the vehicles with the strongest channel gain; and 3) the \emph{joint-importance-and-channel aware} (ICA) scheme \cite{ren2020scheduling}. 

\textbf{Performance}. 
The average precisions, measured by \emph{intersection of union} (IoU) between the prediction and the ground truth, at the communication time of $6000\,$s and $14000\,$s are shown in Fig.~2a,~2b, respectively.
It is observed from all figures that the CA scheme leads to the worst precision at vehicle $1$ whose channel fading is the largest, and the IA scheme leads to the worst precision at vehicle $4$ whose data is less important than others. 
The ICA scheme aims to find a balance between CA and IA, but fails to compete with the proposed scheme due to its heuristic nature as mentioned in the introduction. 
Furthermore, the proposed CTM achieves slightly better performance at the early training stage, but significantly outperforms all the other schemes after sufficient training. 
{This is due to the superiority of \emph{look-ahead nature} inherent in the proposed CTM scheme over the \emph{myopic nature} in the existing CA, IA, and ICA schemes.}
The detection result of one data frame is also illustrated in Fig.~3. 
The red box and green box denote the ground truth and the predicted result, respectively. 
It can be seen that the CTM successfully detect two objects while other schemes only detect one of them. 
Moreover, the CTM achieves the largest IoU with the same time budget.

\vspace{-2mm}
\section{Concluding Remarks}\label{sec:Concluding remarks}
\vspace{-1mm}	
In this paper, we made the first attempt to directly formulate and solve the optimal device scheduling problem in FEEL for communication time minimization. In contrast with the existing solution that can only partially tackle the problem by considering either communication rounds and per-round latency, the proposed solution can fully account for both metrics in the formulated optimization problem. The derived optimized solution shows that it is desired to put more weight in suppressing the remaining communication rounds at the early training stage while gradually bias for reducing per-round latency in the latter  stage. For future work, it is interesting to extend the current work to the model-averaging based federated learning where model updates instead of gradient updates are transmitted. 

\vspace{-2mm}
\section{Acknowledgement}
\vspace{-1mm}
The work was supported in part by the Key Area R\&D Program of Guangdong Province with grant No. 2018B030338001, by the National Key R\&D Program of China with grant No. 2018YFB1800800, by National Natural
Science Foundation of China (No. 62001310), by Shenzhen Outstanding Talents Training Fund, and by Guangdong Research Project No. 2017ZT07X152.


\vspace{-2mm}
\appendix
\vspace{-1mm}
	\subsection{Proof of Theorem \ref{theo:communication upper bound}
} \label{app:theo:upper bound}
	To start with, the following lemmas are needed.
	\begin{lemma}[\cite{ren2020scheduling}]\label{theo:upper bound of loss function}
		\emph{With Assumptions 1 and 2, the upper bound of loss function is given by:}
		\begin{align}
			\mathbb{E} 	\left\{L\left(\mathbf{w}^{\left({N_{t+1}^{\mathbb{E}}+t+1}\right)}\right)-L\left(\mathbf{w}^{*}\right)\right\} \leq \frac{\zeta^{(t+1)}}{N_{t+1}^{\mathbb{E}}+t+\nu +1},
		\end{align}
	\end{lemma}
\noindent	and $\zeta^{(t+1)} \!=\! \max \! \left\{\frac{\ell  \chi^2 (G^{(t+1)})^2}{2(2\mu\chi-1)},(t\! +\! \nu \! +\! 1) \mathbb{E}\left\{\!L\left(\mathbf{w}^{(t+1)}\right) \!\!-\!\! L\left(\mathbf{w}^{*}\right)\right\} \! \right\}$.  
	\begin{lemma}\label{lemma:model state loss function bound} 
		\emph{After the round-$t$'s update, The expected optimality gap between $\mathbf{w}^{(t+1)}$ with $\mathbf{w}^*$ is bounded by}
		\begin{align}\label{expected loss function value upper bound}
			\mathbb{E}&\left\{{L(\mathbf{w}{^{(t+1)}})-L(\mathbf{w}^*)}\right\}\leq\notag \\
			&\left(\frac{1}{2\mu}-\eta^{(t)}\right)\|\mathbf{g}^{(t)}\|^2+\frac{\ell\left(\eta^{(t)}\right)^2}{2}\sum_{m=1}^{M}{\left(\frac{n_m}{n}\right)^2\frac{\|{\mathbf{g}}_{m}^{(t)}\|^2}{p_{m}^{(t)}}}.
		\end{align}
	\end{lemma}
	Based on the Lemma \ref{theo:upper bound of loss function}, to achieve $\epsilon$-accuracy, we have 
	\begin{align} \label{eq:model state based upper bound of remaining rounds}
	N_{t+1}^{\mathbb{E}} \leq \frac{\zeta^{(t)}}{\epsilon}-t-\nu-1, 
	\end{align}
	as $\zeta^{(t+1)}\leq \frac{\ell  \chi^2 \left(G^{(t+1)}\right)^2}{2(2\mu\chi-1)} + (t+1+\nu) \mathbb{E}\left\{L\left(\mathbf{w}^{(t+1)}\right)-L\left(\mathbf{w}^{*}\right)\right\}$	
	substituting the upper bound of $\zeta^{(t+1)}$ into (\ref{eq:model state based upper bound of remaining rounds}), 
	next further substituting Lemma \ref{lemma:model state loss function bound}, we have:
	\begin{align}
		{N}^{\mathbb{E}}_{t+1}
		\leq \frac{\ell(t+1+\nu)\left(\eta^{(t)}\right)^2}{2\epsilon} \sum_{m=1}^{M}{\left(\frac{n_m}{n}\right)^2\frac{\|{\mathbf{g}}_{m}^{(t)}\|^2}{p_{m}^{(t)}}} + C^{(t+1)}\notag.
	\end{align}	

\vspace{-2mm}
\bibliographystyle{ieeetr}

\bibliography{BibDesk_File_v2}
\vspace{-2mm}

\end{document}

%% file: header.tex
\newtheorem{lemma}{Lemma}

\newtheorem{proposition}{Proposition}

\newtheorem{assumption}{Assumption}

\newtheorem{remark}{\bf Remark}

\def\qed{$\Box$}

\def\proof{\noindent{\emph{Proof:} }}

\def
\endproof{\hspace*{\fill}~\qed
\par
\endtrivlist\unskip}
\def\endproof{\hspace*{\fill}~\qed\par\endtrivlist\vskip3pt}

\def\phi{\varphi}

\def\({\left(}
\def\){\right)}

\def\bw{{\mathbf{w}}}
\def\bx{{\mathbf{x}}}

\def\b0{{\mathbf{0}}}

